\documentclass[column,preprintnumbers,amsmath,amssymb]{revtex4}
\preprint{}
\usepackage{graphicx}
\usepackage{dcolumn}
\usepackage{bm}
\usepackage{mathrsfs}
\begin{document}

\title{Optimal control for multi-parameter quantum estimation
with time-dependent Hamiltonians }

\author{Dong  Xie}
\email{xiedong@mail.ustc.edu.cn}
\affiliation{College of Science, Guilin University of Aerospace Technology, Guilin, Guangxi, P.R. China.}

\author{Chunling Xu}
\affiliation{College of Science, Guilin University of Aerospace Technology, Guilin, Guangxi, P.R. China.}

\begin{abstract}
We investigate simultaneous estimation of multi-parameter quantum estimation
with time-dependent Hamiltonians. We analytically obtain the maximal quantum Fisher information matrix for two-parameter in time-dependent three-level systems. The optimal coherent control scheme is proposed to increase the estimation precisions. In a example of a spin-1 particle in a uniformly rotating magnetic field, the optimal coherent Hamiltonians for different parameters can be chosen to be completely same. However, in general, the optimal coherent Hamiltonians for different parameters are incompatibility. In this situation, we suggest a variance method to obtain the optimal coherent Hamiltonian for estimating multiple parameters simultaneously, and obtain the optimal simultaneous estimation precision of two-parameter in a three-level Landau-Zener Hamiltonian.
\end{abstract}

\maketitle

\section{Introduction}
Quantum estimation mainly concerns obtaining fundamental sensitivity limits and developing strategies to enhance
the precision of parameter estimation by utilizing non-classical quantum resource.
Quantum estimation theory were laid in the sixties and seventies, with the two most significant
contributions from Holevo\cite{lab1} and Helstrom\cite{lab2}.

Since its birth, most works of quantum estimation were focused on a single parameter with time-independent Hamiltonians\cite{lab3,lab4,lab5,lab6,lab7,lab8,lab9,lab10,lab11}. In reality, many factors
that influence the systems are changing with time, for example, periodic driving fields or fluctuating external noise. Therefore, quantum estimation with time-independent Hamiltonians significantly limits application of quantum metrology in broader areas. Recently, Ref.\cite{lab12,lab13,lab141,lab14} has studied quantum estimation with time-dependent Hamiltonians. Specially, Ref.\cite{lab14} obtained the maximum quantum Fisher information for parameters in time-dependent Hamiltonians in general, and showed that it is attainable only with proper control on the Hamiltonians generally.

On the other hand, simultaneous quantum-enhanced estimation of multiple parameters with time-independent Hamiltonians is drawing more and more attention. It is mainly because of the fact that unlike in the quantum single-parameter estimation case, quantum measurements required to attain multi-parameter bounds do not necessarily commute\cite{lab15,lab16}. Multi-parameter estimation also has many important applications, such as, quantum imaging\cite{lab17,lab18,lab19}, microscopy and astronomy\cite{lab20,lab21}, sensor networks\cite{lab22,lab23}. All these tasks go beyond single-parameter estimation. There are a lot of theoretical works\cite{lab24,lab25,lab26,lab27,lab28,lab29,lab30,lab31,lab32,lab33}, which clearly show that simultaneous estimation can be more precise than estimating the parameters individually.

Most existing schemes in multi-parameter quantum estimation have assumed that the Hamiltonian is time-independent and the dynamics is fixed, focusing on the identification of the optimal probe states and the optimal measurements. Additional controls usually can be employed to alter the dynamics for further improvement of the precision limit. Some theoretical works\cite{lab141,lab14,lab34,lab35} have recently employed controls to improve the precision limit for single-parameter quantum estimation.  Ref.\cite{lab351} experimentally realized the quantum
improvement in frequency sensitivity with superconducting circuits, using a single transmon qubit. For multi-parameter quantum estimation, the optimal controls have only been obtained under unitary dynamics with time-independent Hamiltonian\cite{lab36}. Given the fixed measurement, Ref.\cite{lab37} obtained the optimal controls for multi-parameter quantum estimation under non-unitary dynamics(general Markovian dynamics with time-independent Hamiltonian). It is highly desired to obtain the optimal controls for the improvement of the precision limit in multi-parameter quantum estimation with time-dependent Hamiltonian.

In this article, we study the multi-parameter quantum estimation with time-dependent Hamiltonians. In order to obtain the analytical results, we only consider two-parameter estimation with three-level Hamiltonians. Firstly, we study a three-level system in a uniformly rotating magnetic field and obtain that the optimal coherent Hamiltonians for two parameters can be chosen to be completely same. However, in general, the optimal coherent Hamiltonians for different parameters are incompatibility. Then, we suggest a variance method to obtain the optimal coherent Hamiltonian for estimating multiple parameters simultaneous, and obtain the optimal simultaneous estimation precision of two-parameters in a three-level Landau-Zener Hamiltonian.

The rest of this article is arranged as follows. In section II, we briefly introduce multi-parameter quantum metrology and propose a universal way to obtain the optimal coherent Hamiltonian for multi-parameter quantum estimation with time-dependent Hamiltonians. In section III, we utilize the optimal control to obtain the optimal two-parameter precision for a three-level system in the model of a uniformly rotating magnetic field. Then, we utilize a variance method to obtain the optimal coherent Hamiltonian for estimating multiple parameters simultaneous, and obtain the optimal simultaneous estimation precision of two-parameter in a three-level Landau-Zener Hamiltonian in section IV.  A conclusion and outlook are presented in section V.
\section{the optimal controls for Multi-parameter Quantum estimation }

In the multi-parameter problem the estimator variance is
promoted to a covariance matrix $\textmd{Cov}(\vec{\lambda})$ and is bounded by
the inverse of the quantum Fisher information matrix through the multi-parameter quantum Fisher information(QFI) Cram\'{e}r-Rao(CR) bound\cite{lab38,lab39}
\begin{equation}
 \textmd{Cov}(\vec{\lambda})\geq F^{-1}(\vec{\lambda}),
\end{equation}
where $\textmd{Cov}(\vec{\lambda})$ refers to the covariance matrix for a locally
unbiased estimator $\vec{\lambda}=(\widetilde{\lambda}_1,\widetilde{\lambda}_2,...,\widetilde{\lambda}_n)$, $\textmd{Cov}(\vec{\lambda})_{jk}=\langle(\widetilde{\lambda}_j-\lambda_j)(\widetilde{\lambda}_k-\lambda_k)\rangle$.
The QFI matrix $F(\vec{\lambda})$ has
matrix elements,
\begin{eqnarray}
F_{\lambda_i\lambda_j}(\vec{\lambda})=\frac{1}{2}\textmd{tr}[\rho_{\vec{\lambda}}(L_{\lambda_i}L_{ \lambda_j}+L_{\lambda_j}L_{\lambda_i })],
\end{eqnarray}
where the symmetric logarithmic  derivative (SLD) $L_{\lambda_i}$ satisfies the equation $\frac{1}{2}(\rho_{\vec{\lambda}} L_{\lambda_i}+L_{\lambda_i}\rho_{\vec{\lambda}})=\partial\rho_{\vec{\lambda}}/\partial L_{\lambda_i}$.
As a result, the estimation cost is bounded by\cite{lab27}
\begin{equation}
 \textmd{Tr}[G\textmd{Cov}(\vec{\lambda})]\geq \textmd{Tr}[GF^{-1}(\vec{\lambda})],
\end{equation}
where $G$ denotes some positive cost matrix, which
allows us to asymmetrically prioritise the uncertainty cost
of different parameters.
When $L_{\lambda_i}$ corresponding to the different parameters commute, there is no additional difficulty in extracting optimal information from a state on all parameters simultaneously. If they do not commute, however, this does not immediately imply that it is impossible to simultaneously extract information on all parameters with precision matching that of the separate scenario for each. The multi-parameter QFI
CR bound can be saturated provided \cite{lab40}
\begin{equation}
\textmd{tr}[\rho_{\vec{\lambda}}[L_{\lambda_i},L_{\lambda_j}]]=0,
\end{equation}
where not the commutator itself but only its expectation value on the probe state is required to vanish.

For pure initial state $|\psi_0\rangle$, the element of quantum Fisher information matrix can be rewritten as
\begin{eqnarray}
F_{\alpha,\beta}(\vec{\lambda})=4(\langle\psi_0|\{\mathcal{H}_\alpha,\mathcal{H}_\beta\}|\psi_0\rangle/2-\langle\psi_0|\mathcal{H}_\alpha|\psi_0\rangle\langle\psi_0|\mathcal{H}_\beta|\psi_0\rangle),
\end{eqnarray}
where the Hermitian operator $\mathcal{H}_m=i(\partial_m U^\dag) U$, $U$ is dependent on a series of parameters $m$ and $\{.,.\}$ denotes the anti-commutation.
The saturation condition Eq.(4) can also be written as
\begin{equation}
\langle\psi_0|[\mathcal{H}_\alpha,\mathcal{H}_\beta]|\psi_0\rangle=0, \forall \alpha, \beta.
\end{equation}

For a time-dependent Hamiltonian  $H(t)$ with multi-parameter $\{m\}$, the corresponding operator $\mathcal{H}_m$ is described by\cite{lab14}
\begin{equation}
\mathcal{H}_m=\int_0^TU^\dag(0\rightarrow t)\partial_mH(t)U(0\rightarrow t),
\end{equation}
where $U(0\rightarrow t)$ is the unitary evolution under the Hamiltonian $H(t)$ for time $t$.

For a single parameter $g$, the maximal QFI can be obtained by using the optimal Hamiltonian\cite{lab14}
\begin{equation}
{H}_c(t,g)=\sum_kf_k(t)|\psi_k(t)\rangle\langle\psi_k(t)|-H(t,g)+i\sum_k|\partial_t\psi_k(t)\rangle\langle\psi_k(t)|,
\end{equation}
where $f_k(t)$ denotes arbitrary real functions and $|\psi_k(t)\rangle$ represents the $k$th eigenstate of $\partial_gH(t,g)$.

We generalize the case of the single parameter to the multi-parameter. We can classify it into two kinds.
First kind, the optimal controls for different parameters are completely same. This situation can be appeared in some special situations, for example, a three-level system in a uniformly rotating magnetic field. The optimal control for simultaneous multi-parameter estimation can be described like the form in Eq.(8)
\begin{equation}
{H}_c(t)=\sum_kf_k(t)|\psi_k(t)\rangle\langle\psi_k(t)|-H(t)+i\sum_k|\partial_t\psi_k(t)\rangle\langle\psi_k(t)|,
\end{equation}
where $|\psi_k(t)\rangle$ represents the $k$th eigenstate of $\partial_mH(t)$.

Second kind, the optimal controls for different parameter can not different. This situation is very general. The optimal control for simultaneous multi-parameter estimation is becoming difficult to be found. We suggest to use a variational method.
Simply, fixing $f_k(t)=0$, the difference comes from the last term of Eq.(9). We suppose that the control Hamiltonian for multi-parameter is given by
\begin{equation}
{H}_c'(t)=-H(t)+i\sum_{m=1}^n\gamma_{m}(\sum_k|\partial_t\psi_k(t)\rangle\langle\psi_k(t)|),
\end{equation}
where the variational parameters $\gamma_m$ are real. One must note that the saturation condition expressed in Eq.(6) should be satisfied. Namely, there are some constraint conditions about the variational parameters $\gamma_m$.

 We can obtain the optimal variational parameters and the optimal controls for multi-parameter estimation, by using the Lagrange multipliers, which is described by
 \begin{eqnarray}
 &\partial_{\gamma_m} L(\gamma_m,\mu_{\alpha\beta})=0, \textmd{for}\ m=1,2...,n;\\
 &\partial_{\mu_{\alpha\beta}} L(\gamma_m,\mu_{\alpha\beta})=0, \textmd{for}\ \ \beta=1,2...,n-1\ \textmd{and}\ \alpha=\beta+1,...,n;
\end{eqnarray}
where $L(\gamma_m,\mu_{\alpha\beta})$ denotes the objective function
\begin{eqnarray}
L(\gamma_m,\mu_{\alpha\beta})=\textmd{Tr}[GF^{-1}(\vec{\lambda})]+\sum_{\alpha=\beta+1}^n\sum_{\beta=1}^{n-1}\mu_{\alpha\beta}\langle\psi_0|[\mathcal{H}_\alpha,\mathcal{H}_\beta]|\psi_0\rangle.
\end{eqnarray}
With the control Hamiltonian, the corresponding operator $\mathcal{H}_m$ is revised as
\begin{equation}
\mathcal{H}_m(T)=\int_0^TU_{tot}^\dag(0\rightarrow t)\partial_mH(t)U_{tot}(0\rightarrow t),
\end{equation}
where $U_{tot}(0\rightarrow t)$ is the unitary evolution under the total Hamiltonian $i\sum_{m=1}^n\gamma_{m}(\sum_k|\partial_t\psi_k(t)\rangle\langle\psi_k(t)|)$ for time $t$.
\section{a spin-1 particle in the model of a uniformly rotating magnetic field }
W consider a spin-j particle in a uniformly rotating magnetic field, $\mathbf{B}(t)=B(\cos wt\mathbf{e_x}+\sin{wt}\mathbf{e_z})$, where $\mathbf{e_x}$ and $\mathbf{e_z}$ are the unit vectors in the $\hat{x}$ and $\hat{z}$ directions, respectively. In order to estimate the amplitude $B$ and the rotation frequency $\omega$ of the field simultaneously in the pure-state case,  we consider the spin $j=1$.
The interaction Hamiltonian $-\mathbf{J}\cdot\mathbf{B}(t)$ between the particle and the field is described by
\begin{equation}
H(t)=-B(\cos \omega tJ_X+\sin \omega tJ_Z),
\end{equation}
where we assume the magnetic moment of the particle to be 1.

We can obtain the analytical result about the Hermitian operators $\mathcal{H}_B$ and $H_\omega$ (in the Appendix A).
When the interrogation time $T\gg1$,
\begin{eqnarray}
&\mathcal{H}_B=-\frac{BT}{\sqrt{B^2+\omega^2}}J_n,\\
&\mathcal{H}_\omega=-\frac{BT}{\sqrt{B^2+\omega^2}}(\cos\sqrt{B^2+\omega^2}TJ_{n\perp}+\sin\sqrt{B^2+\omega^2}TJ_Z),
\end{eqnarray}
where $J_n=\cos\theta J_x+\sin\theta J_y$, $J_{n\perp}=-\sin\theta J_x+\cos\theta J_y$ with $\cos\theta=\frac{B}{\sqrt{B^2+{\omega}^2}}$.
The optimal simultaneous estimation precision are obtained with the initial state $|\psi_0\rangle=\exp(i\theta J_Z)\exp[i(\frac{\pi}{2}-\sqrt{B^2+{\omega}^2}T)J_X](|-1\rangle_Z+|1\rangle_Z)/\sqrt{2}$,
\begin{eqnarray}
\Delta^2\omega=\Delta^2B=\frac{B^2+\omega^2}{4B^2T^2}.
\end{eqnarray}
\subsection{Optimal control}
Next, we try to obtain the optimal control Hamiltonian.
The eigenstates of $\partial_ BH(t)$ can be derived, $|\psi_k(t)\rangle_B=\exp(i\omega tJ_Y)|k\rangle_X$ with $k=-1,0,1$. The eigenstates of $\partial_ \omega H(t)$ are given by, $|\psi_k(t)\rangle_\omega=\exp(i\omega tJ_Y)|k\rangle_Z.$ So the last term in Eq.(9) for $B$ and $\omega$ are completely same, $\sum_ki|\partial_t\psi_k(t)\rangle_\omega\langle\psi_k(t)|=\sum_ki|\partial_t\psi_k(t)\rangle_B\langle\psi_k(t)|=-\omega J_Y$.
We choose the first term in Eq.(9) to be zero. Therefore, the optimal controls for $B$ and $\omega$ are same,
\begin{eqnarray}
H_c(t)=-H(t)-\omega J_Y.
\end{eqnarray}

With the optimal controls, the Hermitian operators $\mathcal{H}_B$ and $\mathcal{H}_\omega$ can be easily derived from Eq.(14) \begin{eqnarray}
\mathcal{H}_B=-TJ_X;\ \mathcal{H}_\omega=-\frac{B}{2}T^2J_Z.
\end{eqnarray}
The optimal initial state can be time-dependent, $|\psi_0\rangle=\frac{1}{\sqrt{2}}(|-1\rangle_Z+|1\rangle_Z)$.
The optimal precision of simultaneous estimation with the optimal controls can be obtained
\begin{eqnarray}
\Delta^2\omega=\frac{1}{B^2T^4},\ \Delta^2B=\frac{1}{4T^2}.
\end{eqnarray}
Comparing with the result in Eq.(18), we can find the optimal control can improve the simultaneous estimation, especially, for large value of $\omega$. And the optimal control can also improve the estimation precision of $\omega$ from the scaling of $T^{-2}$ to $T^{-4}$.
\subsection{Practical control}
Due to the value of $B$ and $\omega$ are unknown, we can only use estimates of $B$ and $\omega$, say $\omega_c$ and $B_c$, instead of the real values of $\omega$ and $B$ in implementing the control Hamiltonian in Eq.(19),
\begin{eqnarray}
H_c(t)=B_c(\cos \omega_c tJ_X+\sin \omega_c tJ_Z)-\omega_c J_Y.
\end{eqnarray}

As shown in Appendix.B, the simultaneous estimation precisions of $\omega$ and $B$ with the control in Eq.(22) is approximately given by 
\begin{eqnarray}
\Delta^2\omega=\frac{1}{B^{2}T^{4}[1-(3T^2\delta\omega^2/4+T^2\delta B^2/2)+2T^2\delta_B\delta_\omega/3]},\\
\Delta^2B=\frac{1}{4T^2[1-(B^2T^4/20+T^2/3)\delta\omega^2-4T^2\delta B^2/9+2T^2\delta B\delta \omega/3]},
\end{eqnarray}
where deviation values $\delta \omega=\omega_c-\omega, \textmd{and } \delta B=B_c-B $.
From the above equations, we can see that the control can dramatically improve the simultaneous estimation precision only when the deviation values are small.

When the measurement can be performed for multiple rounds, the estimate $B_c$ and $\omega_c$ will approach the real value of $B$ and $\omega$, and the estimation precision can be obtained by adaptively updating the
estimate of $B$ and $\omega$ in the control Hamiltonian.

The simplest feedback control strategy is just to first obtain initial estimates of $\omega$ and $B$ without any Hamiltonian control, then use them in the Hamiltonian control in Eq.(9) to produce a high precision estimate of $\omega$ and $B$.
In detail, one measures $\omega$ and $B$ for $N$ times without extra Hamiltonian control.  According to Eq.(18), the variance of the initial estimate of $\omega$ and $B$ are
\begin{eqnarray}
\langle\delta\omega^2\rangle=\langle\delta B^2\rangle=\frac{B^2+\omega^2}{4NB^2T^2},
\end{eqnarray}
Then we apply the control Hamiltonianin in Eq.(22) with this estimates of $\omega$ and $B$. Let the system evolve for the same time T, and this procedure is also repeated for N times. Considering unbiased estimation $\langle\delta\omega\rangle=\langle\delta B\rangle=0$, the final simultaneous estimation precisions of $\omega$ and $B$ should be
\begin{eqnarray}
\Delta^2\omega=\frac{1}{B^{2}T^{4}[1-\frac{5(B^2+\omega^2)}{16NB^2}]},\\
\Delta^2B=\frac{1}{4T^2[1-\frac{(\frac{7}{9}+\frac{B^2T^2}{20})(B^2+\omega^2)}{4NB^2}]}.
\end{eqnarray}
When $N\gg\frac{(25+B^2T^2)(B^2+\omega^2)}{80B^2}$, the optimal estimation precisions as shown in Eq.(21)  will be obtained.

Obviously, this is not the optimal feedback control scheme. For single parameter estimation, many times updating is useful\cite{lab14}. Hence, many steps feedback controls will also perform better for multi-parameter estimation. It is nontrivial for generalizing the feedback control scheme of a single parameter to the case of multi-parameter. We leave it as an open question.

\section{ three-level Landau-Zener Hamiltonian }
The Landau-Zener (LZ) problem is one of the basic paradigms in the physics of quantum systems under the influence of time-dependent Hamiltonians.
There are a lot of works about the two-level Landau-Zener Hamiltonian\cite{lab41,lab42}, which is extremely valuable in understanding the dynamics of quantum systems at avoided crossings of energy levels. In many realistic problems there are more than two energy levels. And multilevel LZ problem has been studied quite extensively in the literature\cite{lab43,lab44,lab45,lab46}.

We consider a three-level Landau-Zener Hamiltonian
\begin{equation}
H_{LZ}(t)=\nu tJ_Z+\Gamma J_X,
\end{equation}
where $\Gamma$ is the level splitting at the transition time $t=0$ and $\nu$ is the speed of the sweep.
And we assume $\nu$ is proportional to $\Gamma$, $\nu=\xi\Gamma$, with the proportionality factor $\xi$.

Next, we use a control Hamiltonian to improve the simultaneous estimation of $\Gamma$ and $\xi$.
Utilizing Eq.(10), the control Hamiltonian is given by
\begin{eqnarray}
 {H}_c(t)=-H_{LZ}(t)+\gamma\frac{\xi}{1+t^2\xi^2}J_Y,
\end{eqnarray}
where $\gamma$ is the variational parameter. When $\gamma=1$, it represents the optimal control for only estimating the single parameter $\Gamma$. When $\gamma=0$, it represents the optimal control for only estimating the single parameter $\xi$.
Then, we can obtain the Hermitian operators for $\Gamma$ and $\xi$
\begin{eqnarray}
 {H}_\Gamma(T)=\int_0^T\sqrt{\xi^2t^2+1}\cos[(1-\gamma_1)\theta]dtJ_X+\int_0^T\sqrt{\xi^2t^2+1}\sin[(1-\gamma_1)\theta]dtJ_Z,\\
 {H}_\xi(T)=\int_0^T\Gamma t\cos(\gamma_1\theta)dtJ_X+\int_0^T\Gamma t\sin(\gamma_1\theta)dtJ_Z,
\end{eqnarray}
where $\theta=\arccos\frac{1}{\sqrt{1+\xi^2t^2}}$.
When we consider the initial pure state $|\psi_0\rangle=\frac{1}{\sqrt{2}}(|-1\rangle_Z+|1\rangle_Z)$, the saturation condition  can be satisfied independent of $\gamma$.
We can numerically obtain the optimal value of $\gamma$ and the estimation precision by using Eq.(13).
\begin{figure}[h]
\includegraphics[scale=1.0]{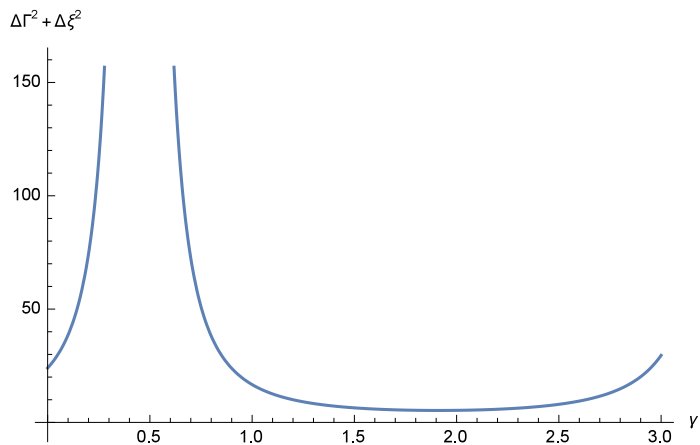}
 \caption{\label{fig.1}Diagram of simultaneous estimation precision. The simultaneous estimation precision $\Delta\gamma^2$ +$\Delta\Gamma^2$ changes with variational parameter $\gamma$. Here, we choose the balanced cost matrix $\mathbf{G}=\mathbf{1}$, $\Gamma=1$, $\xi=1$, and $T=1$.  }
 \end{figure}
 From Fig.1, we can see that the optimal simultaneous estimation precision $\Delta\gamma^2 +\Delta\Gamma^2=5.27421$ can be obtained with the variational value $\gamma=1.9095$. It shows that the optimal control Hamiltonian for simultaneous estimation of $\xi$ and $\Gamma$ is not the optimal control Hamiltonian for the single parameter estimation($\gamma=0$ or $\gamma=1$).
 \section{Conclusion and outlook}
 We have proposed a universal way to obtain the optimal control Hamiltonian for multi-parameter estimation with time-dependent Hamiltonians. For a spin-1 particle in the model of a uniformly rotating magnetic field, we analytical obtain the optimal control Hamiltonian for simultaneous estimation of the amplitude $B$ and the rotation frequency $\omega$ of the field.  The influence of practical control have been shown, and a simple feedback control have been suggested. We use the variational approach to obtain the simultaneous estimation of the splitting level $\Gamma$ and proportional parameter $\xi$ in three-level Landau-Zener Hamiltonian. Numerical results show that, in general, the optimal control Hamiltonian for simultaneous estimation of $\xi$ and $\Gamma$ is not the optimal control Hamiltonian for the single parameter estimation.

Our work is significant for finding the optimal control and obtaining the theoretical limit of multi-parameter estimation precision with time-dependent Hamiltonians. Due to the values of multi-parameter are unknown, we can only use estimates of multi-parameter. The optimal control Hamiltonian can only be obtained by accident. In general, multiple feedback controls is beneficial to improving the multi-parameter estimation precision. Therefore, it deserves to further study on the open question: obtaining a optimized practical feedback control scheme for multi-parameter estimation with time-dependent Hamiltonians.

\section*{Acknowledgement}
 This research was supported by the National
Natural Science Foundation of China under Grant No. 11747008 and Guangxi Natural Science Foundation 2016GXNSFBA380227.
\
\section*{}
$\mathbf{Appendix\ A.}$\\
Utilizing communication relations $[J_X,J_Y]=iJ_Z$, $[J_Y,J_Z]=iJ_X$ and $[J_Z,J_X]=iJ_Y$, we can obtain the useful equations
\begin{eqnarray}
\exp[i\phi J_Y]J_X\exp[-i\phi J_Y]=\cos\phi J_X+\sin\phi J_Z;\ \
\exp[i\phi J_Z]J_X\exp[-i\phi J_Z]=\cos\phi J_X-\sin\phi J_Y.
\end{eqnarray}
Using the above equation, the Hamiltonian in Eq.(15) can be expressed as $H(t)=-\exp(i\omega tJ_Y)J_X\exp(-i\omega tJ_Y)$. Then the unitary operator under $H(t)$ is given by
\begin{eqnarray}
U(0\longrightarrow t)=\exp(i\omega tJ_Y)\exp[i(BJ_X-\omega J_Y)t].
\end{eqnarray}

The Hermitian operators $\mathcal{H}_B$ and $H_\omega$ can be derived
\begin{eqnarray}
&\mathcal{H}_B=\int_0^TU^\dagger(0\longrightarrow t)\partial_BH(t)U(0\longrightarrow t)dt\nonumber\\
&=-\cos\theta J_nT-\frac{\sin\theta\sin(T\sqrt{B^2+\omega^2})}{\sqrt{B^2+\omega^2}}J_{n\perp}+\frac{\sin\theta(1-\cos(T\sqrt{B^2+\omega^2})}{\sqrt{B^2+\omega^2}}J_{Z}.
\end{eqnarray}
\begin{eqnarray}
&\mathcal{H}_\omega=\int_0^TU^\dagger(0\longrightarrow t)\partial_\omega H(t)U(0\longrightarrow t)dt\nonumber\\
&=B(-\frac{T\sin(T\sqrt{B^2+\omega^2})}{\sqrt{B^2+\omega^2}}+\frac{1-\cos(T\sqrt{B^2+\omega^2})}{\sqrt{B^2+\omega^2}})J_Z+B(\frac{\sin(T\sqrt{B^2+\omega^2})}{\sqrt{B^2+\omega^2}}-\frac{T\cos(T\sqrt{B^2+\omega^2})}{\sqrt{B^2+\omega^2}})J_{n\perp}.
\end{eqnarray}

\section*{}
$\mathbf{Appendix\ B.}$\\
As $\omega_c$, $B_c$ are close to $\omega$ and $B$, we expand $\mathcal{H}_B$ ($\mathcal{H}_\omega$) with $\omega_c$ ($B_c$) near $\omega $($B$).

Utilizing Eq.(7), we obtain
\begin{eqnarray}
\mathcal{H}_\omega(T)|_{\omega_c=\omega,B_c=B}=\int_0^T[U^\dag(0\rightarrow t)\partial_\omega H(t)U(0\rightarrow t)]_{\omega_c=\omega,B_c=B}=-\frac{BT^2}{2}J_Z,\\
\mathcal{H}_B(T)|_{\omega_c=\omega,B_c=B}=\int_0^T[U^\dag(0\rightarrow t)\partial_B H(t)U(0\rightarrow t)]_{\omega_c=\omega,B_c=B}=-TJ_X,
\end{eqnarray}
where $U^\dag(0\rightarrow t)$ is the unitary under $H_{tot}=-B(\cos \omega tJ_X+\sin \omega tJ_Z)+B_c(\cos \omega_c tJ_X+\sin \omega_c tJ_Z)-\omega_c J_Y$, $U(0\rightarrow t)|{\omega_c=\omega,B_c=B}=\exp(i\omega J_Y)$.
Also from Eq.(7), we have
\begin{eqnarray}
\partial_{\omega_c}\mathcal{H}_\omega(T)|_{\omega_c=\omega,B_c=B}=\int_0^T[\partial_{\omega_c}U^\dag(0\rightarrow t)\partial_\omega H(t)U(0\rightarrow t)+U^\dag(0\rightarrow t)\partial_\omega H(t)\partial_{\omega_c}U(0\rightarrow t)]_{\omega_c=\omega,B_c=B},\\
\partial_{B_c}\mathcal{H}_\omega(T)|_{\omega_c=\omega,B_c=B}=\int_0^T[\partial_{B_c}U^\dag(0\rightarrow t)\partial_\omega H(t)U(0\rightarrow t)+U^\dag(0\rightarrow t)\partial_\omega H(t)\partial_{B_c}U(0\rightarrow t)]_{\omega_c=\omega,B_c=B},\\
\partial_{\omega_c}\mathcal{H}_B(T)|_{\omega_c=\omega,B_c=B}=\int_0^T[\partial_{\omega_c}U^\dag(0\rightarrow t)\partial_B H(t)U(0\rightarrow t)+U^\dag(0\rightarrow t)\partial_BH(t)\partial_{\omega_c}U(0\rightarrow t)]_{\omega_c=\omega,B_c=B},\\
\partial_{B_c}\mathcal{H}_\omega(T)|_{\omega_c=\omega,B_c=B}=\int_0^T[\partial_{B_c}U^\dag(0\rightarrow t)\partial_B H(t)U(0\rightarrow t)+U^\dag(0\rightarrow t)\partial_B H(t)\partial_{B_c}U(0\rightarrow t)]_{\omega_c=\omega,B_c=B},
\end{eqnarray}
where we have used $\partial_{\omega_c}\partial_\omega H(t)=0$ and $\partial_B\partial_\omega H(t)=0$.
Utilizing a useful equation $\partial_mU(0\rightarrow T)=-i\int_0^TU(t\rightarrow T)\partial_m H_{tot}U(0\rightarrow t)$ as shown in \cite{lab14}, $U(0\rightarrow t)|_{\omega_c=\omega,B_c=B}=\exp(i\omega J_Y)$ and Eq.(28),
we can obtain that
\begin{eqnarray}
\partial_{\omega_c}U(t'\rightarrow t)=-i\exp(i\omega tJ_Y)[\frac{B}{2}(t^2-t'^2)J_Z-J_Y(t-t')]\exp(-i\omega t'J_Y),\\
\partial_{B_c}U(t'\rightarrow t)=-i\exp(i\omega tJ_Y)[\frac{B}{2}J_X(t-t')\exp(-i\omega t'J_Y).
\end{eqnarray}
Substituting the above equations into Eq.(34-37), we can obtain
\begin{eqnarray}
\partial_{\omega_c}\mathcal{H}_\omega(T)|_{\omega_c=\omega,B_c=B}=-\frac{BT^3}{3}J_X,\\
\partial_{B_c}\mathcal{H}_\omega(T)|_{\omega_c,B_c=B}=-\frac{BT^3}{3}J_Y,\\
\partial_{\omega_c}\mathcal{H}_B(T)|_{\omega_c=\omega,B_c=B}=\frac{BT^3}{6}J_Y+\frac{T^2}{2}J_Z,\\
\partial_{B_c}\mathcal{H}_B(T)|_{\omega_c=\omega,B_c=B}=0.
\end{eqnarray}

Similarly, we obtain
\begin{eqnarray}
&\partial^2_{\omega_c}U(0\rightarrow t)_{\omega_c=\omega,B_c=B}=-\exp(i\omega tJ_Y)J_X^2t^2,\\
&\partial^2_{B_c}U(0\rightarrow t)_{\omega_c=\omega,B_c=B}=-\exp(i\omega tJ_Y)(B^2t^4J_Z^2/4-Bt^3J_YJ_Z/3-2Bt^3J_ZJ_Y/3+t^2J_Y^2-iBt^3J_X/3),\\
&\partial_{\omega_c}\partial_{B_c}\mathcal{H}_B(T)|_{\omega_c=\omega,B_c=B}=-\exp(i\omega tJ_Y)[it^2J_Z/2+Bt^3J_XJ_Z/6+Bt^3J_ZJ_X/3-t^2(J_XJ_Y+J_YJ_Z)/2].
\end{eqnarray}
Then, it can be obtained that
\begin{eqnarray}
&\partial^2_{\omega_c}\mathcal{H}_B(T)|_{\omega_c=\omega,B_c=B}=(\frac{B^2T^5}{20}+\frac{T^3}{3})J_X,\\
&\partial^2_{B_c}\mathcal{H}_\omega(T)|_{\omega_c,B_c=B}=\frac{BT^4J_Z}{4},\\
&\partial^2_{\omega_c}\mathcal{H}_\omega(T)|_{\omega_c=\omega,B_c=B}=(\frac{2B^2T^5}{15}J_Y+\frac{BT^4}{4}J_Z),\\
&\partial^2_{B_c}\mathcal{H}_B(T)|_{\omega_c,B_c=B}=0,\\
&\partial_{\omega_c}\partial_{B_c}\mathcal{H}_B(T)|_{\omega_c=\omega,B_c=B}=\frac{T^3}{3}J_Y,\\
&\partial_{\omega_c}\partial_{B_c}\mathcal{H}_\omega(T)|_{\omega_c=\omega,B_c=B}=-\frac{B^2T^5}{30}J_X.
\end{eqnarray}

Therefore, $\mathcal{H}_B(T)$ and $\mathcal{H}_\omega(T)$ can be expand in the vicinity of $\omega=\omega_c$ and $B=B_c$ as
\begin{eqnarray}
&\mathcal{H}_\omega(T)=-\frac{BT^2}{2}J_Z-BT^3J_X\delta B/3-BT^3J_Y\delta\omega/3+(B^2T^5J_Y/15+BT^4J_Z/8)\delta\omega^2\nonumber\\
&+BT^4J_Z\delta B^2/8-B^2T^5J_X\delta B\delta\omega/30+O(\delta\omega^3)+O(\delta B^3)+O(\delta B\delta\omega^2)+O(\delta B^2\delta\omega),\\
&\mathcal{H}_B(T)=-TJ_X+(BT^3J_Y/6+T^2J_Z/2)\delta\omega+\delta B\delta\omega\nonumber\\
&+(B^2T^5/40+T^3/6)\delta \omega^2+O(\delta\omega^3)+O(\delta B^3)+O(\delta B\delta\omega^2)+O(\delta B^2\delta\omega).
\end{eqnarray}

The quantum Fisher matrix can be obtained by Eq.(5), given the initial pure state $|\psi_0\rangle=\frac{1}{\sqrt{2}}(|-1\rangle_Z+|1\rangle_Z)$.
\begin{eqnarray}
F_{BB}=4T^2[1-(B^2T^3/20+T^2/12)\delta\omega^2],\\
F_{\omega \omega}=B^2T^4[1-(T^2\delta\omega^2/2+T^2\delta B^2/18)],\\
F_{\omega B}=F_{B\omega}=-BT^4\delta\omega+4BT^4\delta B/3+2B^2T^5\delta B\delta\omega/15,
\end{eqnarray}
where $\delta B=B_c-B$, $\delta \omega=\omega_c-\omega$.
Simple calculation shows that $\langle\psi_0|[\mathcal{H}_\omega(T),\mathcal{H}_B(T)]|\psi_0\rangle=0$. Hence, the multi-parameter QFI
CR bound can be saturated. The simultaneous estimation precisions are
\begin{eqnarray}
\Delta B^2=\frac{F_{\omega \omega}}{F_{\omega \omega}F_{BB}-F_{\omega B}^2}\approx\frac{1}{4T^2[1-(B^2T^4/20+T^2/3)\delta\omega^2-4T^2\delta B^2/9+2T^2\delta B\delta \omega/3]},\\
\Delta \omega^2=\frac{F_{BB}}{F_{\omega \omega}F_{BB}-F_{\omega B}^2}\approx\frac{1}{B^{2}T^{4}[1-(3T^2\delta\omega^2/4+T^2\delta B^2/2)+2T^2\delta_B\delta_\omega/3]}.
\end{eqnarray}

\end{document}